\documentclass[letterpaper, 10 pt, conference]{ieeeconf}
\IEEEoverridecommandlockouts
\usepackage{graphics} 
\usepackage{epsfig} 
\usepackage{mathrsfs}
\usepackage{times} 
\usepackage{amsmath} 
\usepackage{amssymb}  
\usepackage{graphicx}
\usepackage{tabularx,array,ragged2e,booktabs}
\usepackage{color}
\usepackage{lipsum}
\usepackage{subcaption}
\usepackage{booktabs}
\usepackage{xcolor}
\usepackage{pifont}
\usepackage{float}

\usepackage{tikz}
\usepackage{pgfplots}
\pgfplotsset{compat=1.18}

\usepackage{booktabs}
\usepackage{siunitx}
\usepackage[style=ieee,dashed=false]{biblatex}

\title{\LARGE \bf
Trajectory-Regularized Stochastic Optimal Control via KL Divergence
}

\author{Mintae Kim and Koushil Sreenath
\thanks{The authors are with \textit{Hybrid Robotics}, UC Berkeley.
This work was supported in part by Design of Robustly Implementable Autonomous and Intelligent Machines, DARPA award number HR00112490425.}
}

\begin{document}

\maketitle
\thispagestyle{empty}
\pagestyle{empty}


\begin{abstract}

We introduce trajectory-regularized stochastic optimal control (TRSOC), which augments standard stochastic optimal control (SOC) with a Kullback--Leibler (KL) divergence between controlled and reference trajectory distributions.
Using Girsanov's theorem, the trajectory KL reduces to a quadratic drift mismatch penalty, yielding a modified running cost that preserves the dynamic programming (DP) structure.
We derive the corresponding Hamilton--Jacobi--Bellman (HJB) equation and characterize the optimal policy.
In the linear-quadratic (LQ) setting, the formulation admits a closed-form solution with an augmented control cost.
Experiments show that the regularization parameter induces a trade-off between performance-driven and reference-preserving behavior, including cases with reference dynamics learned from offline data.

\end{abstract}

\begingroup
\renewcommand{\thefootnote}{}
\footnotetext{
Email: \texttt{\{mintae.kim, koushils\}@berkeley.edu}
}
\endgroup


\section{Introduction}

\subsection{Motivation}

Stochastic optimal control (SOC) provides a framework for controlling systems under uncertainty \cite{fleming2006controlled, yong1999stochastic, kim2026finite}.
In the classical setting, an admissible control minimizes an expected cumulative cost under stochastic dynamics.
Optimal policies are characterized by dynamic programming (DP) and Hamilton--Jacobi--Bellman (HJB) equations \cite{bertsekas2012dynamic}.
However, standard SOC focuses primarily on task performance.
It does not explicitly control how far the resulting dynamics deviate from a desired reference behavior.
In many applications, such reference behaviors are available and encode useful structure, including passive dynamics, baseline closed-loop dynamics, and reference dynamics estimated from offline trajectories.
Large deviations from such references may discard useful prior structure or produce behavior far from an available nominal process.
This motivates a behavior-regularized SOC formulation that balances performance and proximity to a reference trajectory distribution.

We measure this deviation using Kullback--Leibler (KL) divergence on trajectory space \cite{leonard2013survey}.
While KL regularization is often applied locally at the action level in reinforcement learning (RL), here we regularize the trajectory distribution induced by the controlled process \cite{schulman2015trust, kim2026robust}.
For controlled diffusions with shared nondegenerate noise, Girsanov's theorem converts this trajectory-level KL into a quadratic penalty on the drift mismatch \cite{karatzas2014brownian}.
This leads to trajectory-regularized stochastic optimal control (TRSOC), which incorporates prior behaviors, including nominal controllers and data-driven references, into the SOC framework.


\subsection{Related Work}

Path-integral (PI) control considers a class of SOC problems in which the HJB equation can be linearized via a log transformation, enabling importance sampling-based solutions such as model predictive PI (MPPI) methods \cite{kappen2007introduction, thijssen2016path, williams2017model, kim2026wombet}.
However, this approach relies on structural matching conditions between the control and diffusion channels \cite{kappen2005linear}.
KL and relative-entropy control formulations penalize deviation from a reference trajectory measure and can yield exponential trajectory reweighting or linearly solvable structure under suitable assumptions \cite{boue1998variational, lehec2013representation, bierkens2014explicit}.
Related ideas also appear in the Schr\"odinger bridge, which seeks an entropy-minimizing interpolation between prescribed endpoint distributions relative to a reference process \cite{leonard2013survey}.

In contrast to KL-control formulations, TRSOC retains the admissible control as the optimization variable and regularizes the trajectory law induced by the controlled SDE.
Using Girsanov's theorem, the trajectory KL admits a local representation as a drift-mismatch penalty, preserving the Markov structure and standard DP framework.
The resulting problem remains within the SOC formulation and retains an HJB characterization for nonlinear controlled diffusions.

\vspace{-0.15em}

\subsection{Contributions}

The main contributions of this paper are as follows:

\begin{itemize}

\item \textbf{TRSOC as behavior-regularized SOC:}
We formulate TRSOC by augmenting the SOC objective with a KL divergence between controlled and reference trajectory measures.

\item \textbf{Trajectory KL to local cost:}
Under shared nondegenerate diffusion, Girsanov's theorem converts the trajectory KL exactly into a quadratic drift-mismatch penalty, preserving the DP and HJB structure and admitting a random-horizon interpretation in the discounted setting.

\item \textbf{Optimal control structure and stability:}
We derive the optimal feedback structure for control-affine systems with quadratic costs, establish sufficient Foster-Lyapunov conditions for closed-loop stability, and characterize the LQ case through a Riccati equation.

\end{itemize}





\section{Problem Formulation}
\label{sec:problem}

\subsection{Controlled and Reference Stochastic Dynamics}

Let $(\Omega,\mathcal F,\{\mathcal F_t\}_{t\ge0},\mathbb P)$ be a filtered probability space.
We consider a controlled diffusion process
\begin{equation}
\label{eq:controlled_sde}
dX_t = f(t,X_t,u_t)dt + \sigma(t,X_t)d\mathcal{W}_t,
\quad
X_0 = x \in \mathbb R^n,
\end{equation}
where $\mathcal{W}_t$ is an $n$-dimensional Wiener process and the control enters through the drift \cite{oksendal2003stochastic}.
The control takes values in a closed convex set $\mathcal U \subset \mathbb R^m$.
We address Markov feedback controls of the form $u_t=u(t,X_t)$.
We denote by $\mathcal A$ the class of such controls taking values in $\mathcal U$ and satisfying
$\mathbb E[\int_0^T \|u(t,X_t)\|^2 dt] < \infty$ for every $T>0$.
We assume that $f$ and $\sigma$ satisfy standard regularity and growth conditions ensuring a unique solution of \eqref{eq:controlled_sde} for every $u\in\mathcal A$ \cite{oksendal2003stochastic}.

To incorporate nominal or prior behavior, we introduce the reference dynamics
\begin{equation}
\label{eq:reference_sde}
dX_t = f_0(t,X_t)dt + \sigma(t,X_t)d\mathcal{W}_t,
\end{equation}
which shares the same diffusion as \eqref{eq:controlled_sde} but has reference drift $f_0$.
This can represent, for example, passive dynamics, a nominal model, closed-loop dynamics induced by a baseline controller, or dynamics estimated from offline trajectory data.
When the reference behavior is generated by a feedback controller $u^0$, the reference drift can be written as $f_0(t,x)=f(t,x,u^0(t,x))$.


\subsection{Trajectory Regularization via KL Divergence}

Fix a finite horizon $T>0$.
For $u \in \mathcal A$, let $X^u=\{X_t^u\}_{t\in[0,T]}$ denote the unique strong solution of the controlled SDE
\eqref{eq:controlled_sde}.
Let $X^0=\{X_t^0\}_{t\in[0,T]}$ denote the corresponding reference process solving
\eqref{eq:reference_sde}.
We write $\mathbb P^u$ and $\mathbb P^0$ for the probability laws of $X^u$ and $X^0$, respectively, on the trajectory space $C([0,T];\mathbb R^n)$.

We impose the following condition for the trajectory-measure comparison and the resulting KL divergence.

\noindent
\textbf{\textit{Assumption 1.}}
For every $(t,x)\in[0,T]\times\mathbb R^n$, the diffusion matrix $\sigma(t,x)$ is invertible.
Moreover, for any $u \in \mathcal A$, define
\begin{equation}
\label{eq:phi_def}
\phi^u(t,x)
:=
\sigma(t,x)^{-1}\bigl(f(t,x,u(t,x))-f_0(t,x)\bigr).
\end{equation}
We assume that Novikov's condition holds under the reference process,
\begin{equation}
\mathbb E^{\mathbb P^0}\left[
\exp\left(\frac12 \int_0^T \|\phi^u(t,X_t^0)\|^2 dt
\right)
\right] < \infty.
\label{eq:novikov_condition}
\end{equation}

Under Assumption~1, Girsanov's theorem guarantees that $\mathbb P^u \ll \mathbb P^0$ on the finite-horizon trajectory space, so the corresponding KL divergence is well defined \cite{girsanov2012lectures}.

Our objective is to regularize deviation from the reference dynamics at the level of trajectory distributions.
A natural measure of such deviation is $\mathrm{KL}(\mathbb P^u \| \mathbb P^0)$.

\noindent
\textbf{\textit{Lemma 1 (Trajectory KL Divergence Identity).}}
Suppose Assumption~1 holds.
Then $\mathbb P^u \ll \mathbb P^0$, and
\begin{equation}
\label{eq:kl_identity}
\mathrm{KL}(\mathbb P^u \| \mathbb P^0)
=
\frac12
\mathbb E^{\mathbb P^u}
\left[
\int_0^T
\|\phi^u(t,X_t^u)\|^2dt
\right].
\end{equation}

\noindent
This follows from Girsanov's theorem under Assumption~1.

Lemma~1 shows that trajectory KL regularization reduces exactly to a quadratic penalty on the drift mismatch between the controlled and reference dynamics.

Although Lemma~1 is stated over $[0,T]$, same holds on any interval $[t,T]$.
More precisely, for any $t \in [0,T]$, the conditional trajectory measures starting from $(t,x)$ satisfy
\begin{equation}
\label{eq:kl_identity_shifted}
\mathrm{KL}\bigl(\mathbb P^u_{t,T} \| \mathbb P^0_{t,T}\bigr)
=
\frac12
\mathbb E_{t,x}^{\mathbb P^u}
\left[
\int_t^T
\|\phi^u(s,X_s^u)\|^2ds
\right],
\end{equation}
where $\mathbb P^u_{t,T}$ and $\mathbb P^0_{t,T}$ denote the conditional laws of the processes over $[t,T]$.
This follows from the Markov property and the time-shifted form of Girsanov's theorem \cite{oksendal2003stochastic}.
Thus, the KL divergence is additive over time, with the same local drift-mismatch contribution as in \eqref{eq:kl_identity}.

\noindent
\textbf{\textit{Remark 1.}}
The representations \eqref{eq:kl_identity} and \eqref{eq:kl_identity_shifted} rely on drift-only control and absolute continuity through Girsanov's theorem \cite{karatzas2014brownian}.
If the diffusion is control-dependent, absolute continuity between the trajectory measures may fail, and the quadratic drift-energy representation need not hold.
In such cases, the trajectory-level KL divergence may no longer admit the simple local form above.


\subsection{Trajectory Regularized Stochastic Optimal Control}

We now define the TRSOC problem.
Let $\ell:[0,T]\times\mathbb R^n \times \mathcal U \to \mathbb R$ be a continuous running cost and let $g:\mathbb R^n \to \mathbb R$ be a continuous terminal cost, both with at most polynomial growth.
For $\lambda \ge 0$, define the KL-augmented running cost
\begin{equation}
\label{eq:augmented_running_cost}
\begin{aligned}
L(t,x,u)
:=
&\ell(t,x,u)
\\
&+\frac{\lambda}{2}
\left\|
\sigma(t,x)^{-1}\bigl(f(t,x,u)-f_0(t,x)\bigr)
\right\|^2.
\end{aligned}
\end{equation}
By the time-shifted representation \eqref{eq:kl_identity_shifted}, the trajectory-level KL divergence over $[t,T]$ admits an additive decomposition.
Thus, it can be incorporated as a running cost through \eqref{eq:augmented_running_cost}.

For a finite $T>0$, initial time $t \in [0,T]$, and initial condition $X_t=x$, the finite-horizon cost of $u \in \mathcal A$ is
\begin{equation}
\label{eq:finite_objective}
J_{t,x}(u)
:=
\mathbb E_{t,x}^{\mathbb P^u}
\left[
\int_t^T L(s,X_s^u,u(s,X_s^u))ds + g(X_T^u)
\right].
\end{equation}
The corresponding value function is
\begin{equation}
\label{eq:finite_value}
V(t,x)
:=
\inf_{u \in \mathcal A} J_{t,x}(u).
\end{equation}

By Lemma~1, the quadratic term in \eqref{eq:augmented_running_cost} is the local form of the trajectory-level KL regularization.
Accordingly, the finite-horizon problem \eqref{eq:finite_objective} balances task performance against deviation from the reference trajectory distribution.

For the stationary infinite-horizon analysis, we assume that $f$, $f_0$, $\sigma$, and $\ell$ are time-homogeneous, that $\ell(x,u)\ge0$, and that Assumption~1 holds on every finite horizon.
We define
$\mathcal A_\rho := \{u\in\mathcal A:
\mathbb E[\int_0^\infty e^{-\rho t}\|u(t,X_t)\|^2dt]<\infty\}$
for discount rate $\rho>0$.
Under this stationary specialization, we write \eqref{eq:augmented_running_cost} as $L(x,u)$, and for $u \in \mathcal A_\rho$ define
\begin{equation}
\label{eq:infinite_objective}
J_x(u)
:=
\mathbb E_x^{\mathbb P^u}
\left[
\int_0^\infty
e^{-\rho t}
L(X_t^u,u(t,X_t^u))dt
\right].
\end{equation}
The associated infinite-horizon value function is defined by
\begin{equation}
\label{eq:infinite_value}
V(x)
:=
\inf_{u \in \mathcal A_\rho} J_x(u).
\end{equation}
With a slight abuse of notation, we write $V$ for the infinite-horizon value function when no ambiguity arises.

Importantly, TRSOC does not require the controlled dynamics to exactly reproduce the reference.
In particular, we do not assume that the reference drift is exactly attainable by the control.
Instead, the reference enters as a soft trajectory-level penalty, yielding a trade-off between task performance and reference deviation.
When exact matching is infeasible, the optimization remains well defined and balances task performance against the achievable deviation from the reference.


\section{Dynamic Programming and HJB Equations}
\label{sec:dp-hjb}

\subsection{Value Function and DP}

Recall from Sec.~\ref{sec:problem} the finite-horizon cost functional \eqref{eq:finite_objective}, the value function \eqref{eq:finite_value}, and the KL-augmented running cost \eqref{eq:augmented_running_cost}.
Here, we work under the assumptions of Sec.~\ref{sec:problem} to derive the DP principle and the corresponding HJB equation.

By Lemma~1 and \eqref{eq:kl_identity_shifted}, the trajectory KL admits an additive running-cost representation over every subinterval.
Hence, the regularized objective remains time-consistent and the standard DP principle applies without state augmentation \cite{fleming2006controlled}.

\noindent
\textbf{\textit{Theorem 1 (Finite-Horizon DP).}}
For all $(t,x)\in[0,T]\times\mathbb R^n$ and any deterministic $h\in[0,T-t]$, $V(t,x)$ satisfies
\begin{equation}
\label{eq:dp_finite}
\begin{aligned}
V(t,x)
=
\inf_{u\in\mathcal A}
\mathbb E_{t,x}^{\mathbb P^u}
\Bigg[
&\int_t^{t+h} L(s,X_s^u,u(s,X_s^u))\,ds
\\
&+V(t+h,X_{t+h}^u)
\Bigg].
\end{aligned}
\end{equation}
The class of time-dependent Markov feedback controls is closed under deterministic-time concatenation, so \eqref{eq:dp_finite} follows from the standard DP principle for controlled diffusions.


\subsection{Finite-Horizon HJB Equation and Viscosity}

We characterize the value function via an HJB equation.
For $\varphi\in C^{1,2}([0,T]\times\mathbb R^n)$ and $u\in\mathcal U$, define the time-dependent generator
\begin{equation}
\label{eq:generator_finite}
\begin{aligned}
\mathcal L_t^u \varphi(t,x)
=
&\nabla_x \varphi(t,x)^\top f(t,x,u)
\\
& +\frac12
\mathrm{tr}\!\left(\sigma(t,x)\sigma(t,x)^\top \nabla_x^2 \varphi(t,x)\right).
\end{aligned}
\end{equation}
By DP (Theorem~1) and It\^o's formula, the value function satisfies the HJB equation \cite{oksendal2003stochastic}
\begin{equation}
\label{eq:hjb_finite}
-\partial_t V(t,x)
=
\inf_{u\in\mathcal U}
\left\{
L(t,x,u) + \mathcal L_t^u V(t,x)
\right\}, \;
V(T,x)=g(x).
\end{equation}
KL regularization enters \eqref{eq:hjb_finite} only through the augmented running cost $L(t,x,u)$, thus preserving the standard DP.

Since $V$ is not necessarily smooth, we interpret \eqref{eq:hjb_finite} in the viscosity sense.
Under standard assumptions ensuring valid DP and the corresponding comparison principle, $V$ can be characterized as the unique viscosity solution of \eqref{eq:hjb_finite} \cite{fleming2006controlled}.

We state a verification result, which provides an optimality guarantee when there exists a smooth solution to \eqref{eq:hjb_finite}.

\noindent
\textbf{\textit{Theorem 2 (Verification Theorem).}}
Let $W:[0,T]\times\mathbb R^n\to\mathbb R$ be a function satisfying:
\begin{itemize}
    \item[(i)] $W \in C^{1,2}([0,T)\times\mathbb R^n) \cap C([0,T]\times\mathbb R^n)$,
    \item[(ii)] $W(T,x)=g(x)$,
    \item[(iii)] $W$ satisfies \eqref{eq:hjb_finite}.
\end{itemize}
Then $W(t,x)\le V(t,x)$ for all $(t,x)$.
If there exists a minimizer $u^*(t,x)$ in \eqref{eq:hjb_finite} such that the induced feedback control is admissible, then $u^*$ is optimal and $W=V$ \cite{yong1999stochastic}.


\subsection{Infinite-Horizon DP and Stationary HJB Equation}

We next consider DP for the stationary discounted infinite-horizon formulation introduced in \eqref{eq:infinite_objective}.

\noindent
\textbf{\textit{Proposition 2 (Infinite-Horizon DP).}}
For all $x\in\mathbb R^n$ and deterministic $h\ge0$,
\begin{equation}
\label{eq:dp_infinite}
\begin{aligned}
V(x)
=
\inf_{u\in\mathcal A_\rho}
\mathbb E_x^{\mathbb P^u}
\Bigg[
&\int_0^h e^{-\rho t} L(X_t^u,u(t,X_t^u))\,dt
\\
&+e^{-\rho h} V(X_h^u)
\Bigg].
\end{aligned}
\end{equation}

For $\varphi\in C^2(\mathbb R^n)$ and $u\in\mathcal U$, define the generator
\begin{equation}
\label{eq:generator_stationary}
\mathcal L^u\varphi(x)
=
\nabla\varphi(x)^\top f(x,u)
+
\frac12
\mathrm{tr}\!\left(\sigma(x)\sigma(x)^\top\nabla^2\varphi(x)\right).
\end{equation}
The corresponding stationary HJB equation is
\begin{equation}
\label{eq:hjb_stationary}
\rho V(x)
=
\inf_{u\in\mathcal U}
\left\{
L(x,u) + \mathcal L^u V(x)
\right\}.
\end{equation}


\section{Structural Properties of TRSOC}
\label{sec:structure}

\subsection{Dependence on the Regularization Parameter}

Throughout this section, we work in the stationary discounted setting and write
$\phi_t^u:=\sigma(X_t^u)^{-1}\bigl(f(X_t^u,u(t,X_t^u))-f_0(X_t^u)\bigr)$.
The regularization parameter $\lambda \ge 0$ governs the trade-off between task performance and proximity to the reference behavior.

\noindent
\textbf{\textit{Proposition 3 (Monotonicity in $\lambda$).}}
Let $V^\lambda$ denote the value function associated with $\lambda \ge 0$.
Then for all $x \in \mathbb R^n$,
\begin{equation}
\label{eq:monotonicity_lambda}
\lambda_1 \le \lambda_2
\quad \Longrightarrow \quad
V^{\lambda_1}(x) \le V^{\lambda_2}(x).
\end{equation}

\noindent
\textit{Proof.}
For any $u\in\mathcal{A}_\rho$,
\[
J_x^{\lambda_2}(u)
=
J_x^{\lambda_1}(u)
+
\frac{\lambda_2-\lambda_1}{2}
\mathbb E_x^{\mathbb P^u}
\int_0^\infty
e^{-\rho t}
\|\phi_t^u\|^2 dt
\ge
J_x^{\lambda_1}(u).
\]
Taking the infimum with respect to $u$ yields the result. $\square$

Thus, increasing $\lambda$ cannot decrease the optimal regularized cost, since it increases the penalty assigned to reference deviation.

Next, consider the limiting regimes.
As $\lambda \to 0$, TRSOC reduces to the standard SOC problem.
If an optimal unregularized control $u^0\in\mathcal A_\rho$ has finite discounted drift deviation, comparison with $u^0$ gives $V^\lambda(x)\to V^0(x)$ as $\lambda\to0$.
The opposite regime $\lambda\to\infty$ induces alignment with the reference whenever exact drift matching is feasible.

\noindent
\textbf{\textit{Proposition 4 (Reference-Preserving Limit).}}
Suppose there exists $\bar u \in \mathcal{A}_\rho$ such that $f(x,\bar u(x)) = f_0(x)$ for all $x\in\mathbb R^n$ and $J_x^0(\bar u)<\infty$.
Let $u^\lambda$ be an optimal control for TRSOC with $\lambda$.
Then, for any $T>0$,
\begin{equation}
\label{eq:lambda_large_drift}
\mathbb E_x^{\mathbb P^{u^\lambda}}
\left[
\int_0^T
\left\|
\phi_t^{u^\lambda}
\right\|^2 dt
\right]
\to 0
\quad
\text{as } \lambda\to\infty.
\end{equation}

\noindent
\textit{Proof.}
Since $u^\lambda$ is optimal and the regularization term vanishes for $\bar u$,
\begin{multline*}
J_x^0(u^\lambda)
+
\frac{\lambda}{2}
\mathbb E_x^{\mathbb P^{u^\lambda}}
\left[
\int_0^\infty
e^{-\rho t}
\left\|
\phi_t^{u^\lambda}
\right\|^2 dt
\right]\\
=
J_x^\lambda(u^\lambda)
\le
J_x^\lambda(\bar u)
=
J_x^0(\bar u).
\end{multline*}
Since the stationary running cost is nonnegative, $J_x^0(u^\lambda)\ge0$.
Using $e^{-\rho t}\ge e^{-\rho T}$ on $[0,T]$ gives
\[
\frac{\lambda}{2} e^{-\rho T}
\mathbb E_x^{\mathbb P^{u^\lambda}}
\left[
\int_0^T
\left\|
\phi_t^{u^\lambda}
\right\|^2 dt
\right]
\le
J_x^0(\bar u).
\]
Hence
\begin{equation*}
\mathbb E_x^{\mathbb P^{u^\lambda}}
\left[
\int_0^T
\left\|
\phi_t^{u^\lambda}
\right\|^2 dt
\right]
\le
\frac{2e^{\rho T}}{\lambda}\,J_x^0(\bar u)
\xrightarrow{\lambda\to\infty} 0.
\quad\square
\end{equation*}

Proposition~4 shows that increasing \(\lambda\) drives the controlled drift toward the reference when exact matching is feasible.


\subsection{Performance--Deviation Tradeoff}

We now quantify the trade-off between task performance and deviation from the reference behavior.

For any $u\in\mathcal{A}_\rho$, the TRSOC objective decomposes as
\begin{equation}
\label{eq:decomposition_objective}
J_x^\lambda(u)
=
J_x^0(u)
+
\frac{\lambda}{2}
\mathbb E_x^{\mathbb P^u}
\left[
\int_0^\infty
e^{-\rho t}
\left\| \phi_t^u \right\|^2 dt
\right].
\end{equation}
The discounted deviation term is not itself the KL divergence between infinite-horizon trajectory measures.
If $\tau\sim\mathrm{Exp}(\rho)$ is independent of the process, Lemma~1 and Tonelli's theorem identify its half-integral with
$\mathbb E_\tau[\mathrm{KL}(\mathbb P^u_{0,\tau}\|\mathbb P^0_{0,\tau})]$.

Let $u^\lambda$ be optimal for TRSOC and $u^0$ optimal for the SOC without trajectory regularization.
Optimality of $u^\lambda$ implies $J_x^\lambda(u^\lambda) \le J_x^\lambda(u^0)$.
Using \eqref{eq:decomposition_objective}, we obtain
\begin{equation}
\label{eq:trade-off_inequality}
0
\le
J_x^0(u^\lambda) - J_x^0(u^0)
\le
\frac{\lambda}{2}
\mathbb E_x^{\mathbb P^{u^0}}
\left[
\int_0^\infty
e^{-\rho t}
\left\|
\phi_t^{u^0}
\right\|^2 dt
\right].
\end{equation}

Inequality \eqref{eq:trade-off_inequality} quantifies the task-performance degradation induced by regularization.
In particular, the suboptimality of $u^\lambda$ relative to the unregularized optimum is controlled by how much $u^0$ deviates from the reference.

This reveals a fundamental trade-off structure: minimizing $J_x^\lambda$ balances the standard performance objective $J_x^0$ against the discounted trajectory-deviation penalty.
Thus, TRSOC can be interpreted as a weighted scalarization of the objectives
\[
J_x^0(u)
\quad\text{vs.}\quad
\mathbb E_x^{\mathbb P^u}
\left[
\int_0^\infty
e^{-\rho t}
\left\|
\phi_t^u
\right\|^2 dt
\right].
\]

Varying $\lambda$ generates supported performance--deviation trade-offs.
As with weighted scalarization generally, it need not recover every Pareto-efficient point when the attainable objective set is nonconvex.


\subsection{Relation to KL Control and Risk-Sensitive Control}

Trajectory-level relative entropy also appears in KL-control formulations in which the trajectory measure is the optimization variable \cite{bierkens2014explicit},
\begin{equation}
\inf_{\mathbb P}
\left\{
\mathbb E^{\mathbb P}\!\left[\int_0^T \ell(X_t)dt\right]
+
\lambda \mathrm{KL}(\mathbb P \| \mathbb P^0)
\right\}.
\end{equation}
This leads to exponential tilting of trajectories and, under suitable structural assumptions, to linearly solvable or entropic control formulations \cite{todorov2006linearly}.

In contrast, TRSOC retains the admissible control as the optimization variable, with the trajectory measure induced by the controlled SDE.
Under Assumption~1, Girsanov's theorem converts the induced finite-horizon trajectory KL exactly into a local drift-mismatch running cost.
Thus, TRSOC remains a standard SOC problem with an augmented cost, preserving the Markov state, DP principle, and HJB equation.
The task objective remains a linear expectation under the controlled process, while the trajectory regularizer penalizes deviation from the reference dynamics.


\section{Structure of the Optimal Control}
\label{sec:optimal}

In this section, we specialize the stationary discounted HJB equation \eqref{eq:hjb_stationary} to control-affine dynamics and quadratic control costs.
We assume $\mathcal U=\mathbb R^m$, so that the control is unconstrained, and derive an explicit optimal feedback law and the corresponding closed-loop dynamics \cite{bertsekas2012dynamic}.
The same argument applies to the finite-horizon HJB equation \eqref{eq:hjb_finite} with time-dependent coefficients and value function $V^\lambda(t,x)$.


\subsection{Control-Affine Drift and Quadratic Control Cost}

Suppose that the drift is control-affine:
\begin{equation}
\label{eq:affine_drift}
f(x,u)=f_0(x)+B(x)u,
\end{equation}
where $B:\mathbb R^n\to\mathbb R^{n\times m}$ satisfies the regularity conditions required for well-posedness.
Thus, $u=0$ reproduces the reference drift.

Under Assumption~1, $\sigma(x)$ is invertible for every $x\in\mathbb R^n$.
Hence the KL regularization term in \eqref{eq:augmented_running_cost} becomes
\begin{equation}
\label{eq:kl_affine}
\frac{\lambda}{2}
\left\|
\sigma(x)^{-1}B(x)u
\right\|^2
=
\frac{\lambda}{2}
u^\top B(x)^\top\bigl(\sigma(x)\sigma(x)^\top\bigr)^{-1}B(x)u.
\end{equation}

Next, suppose that the running cost has the form
\begin{equation}
\label{eq:quadratic_running_cost}
\ell(x,u)=q(x)+\frac12 u^\top R(x)u,
\end{equation}
where $q:\mathbb R^n\to\mathbb R_{\ge0}$ is continuous and $R(x)\in\mathbb R^{m\times m}$ is positive definite for all $x\in\mathbb R^n$.

Define the $\lambda$-dependent effective control weight
\begin{equation}
\label{eq:effective_control_weight}
\widetilde R_\lambda(x)
:=
R(x)
+
\lambda B(x)^\top\bigl(\sigma(x)\sigma(x)^\top\bigr)^{-1}B(x).
\end{equation}
Since $R(x)\succ0$ and $\lambda\ge0$, we have $\widetilde R_\lambda(x)\succ0$.

Then, the stationary HJB equation \eqref{eq:hjb_stationary} becomes
\begin{multline}
\label{eq:hjb_hamiltonian}
\rho V^\lambda(x)
=
\inf_{u\in\mathbb R^m}
\Bigl\{
q(x)
+
\frac12 u^\top \widetilde R_\lambda(x)u
+
\nabla V^\lambda(x)^\top f_0(x)
\\
+
\nabla V^\lambda(x)^\top B(x)u
+
\frac12
\mathrm{tr}\!\left(\sigma(x)\sigma(x)^\top \nabla^2V^\lambda(x)\right)
\Bigr\}.
\end{multline}


\subsection{Explicit Optimal Feedback Law}

The minimization in \eqref{eq:hjb_hamiltonian} is strictly convex in $u$, so its pointwise minimizer can be computed explicitly.

\noindent
\textbf{\textit{Proposition 5 (Explicit Optimal Feedback Law).}}
Assume \eqref{eq:affine_drift} and \eqref{eq:quadratic_running_cost}.
Then the unique minimizer of \eqref{eq:hjb_hamiltonian} is
\begin{equation}
\label{eq:optimal_feedback_stationary}
u^{*,\lambda}(x)
=
-\widetilde R_\lambda(x)^{-1}B(x)^\top \nabla V^\lambda(x).
\end{equation}
If the resulting feedback is admissible, it is optimal by the verification theorem.

\noindent
\textit{Proof.}
Differentiating the RHS of \eqref{eq:hjb_hamiltonian} with respect to $u$ and setting the derivative equal to zero yields
$\widetilde R_\lambda(x)u+B(x)^\top\nabla V^\lambda(x)=0$.
Since $\widetilde R_\lambda(x)\succ0$, the Hamiltonian is convex in $u$, and the unique minimizer is given by \eqref{eq:optimal_feedback_stationary}. $\square$

Substituting \eqref{eq:optimal_feedback_stationary} back into \eqref{eq:hjb_hamiltonian} gives the reduced stationary HJB equation
\begin{multline}
\label{eq:hjb_reduced}
\rho V^\lambda(x)
=
q(x)
+
\nabla V^\lambda(x)^\top f_0(x)
+
\frac12
\mathrm{tr}\!\left(\sigma(x)\sigma(x)^\top \nabla^2V^\lambda(x)\right)
\\
-
\frac12
\nabla V^\lambda(x)^\top
B(x)\widetilde R_\lambda(x)^{-1}B(x)^\top
\nabla V^\lambda(x).
\end{multline}

The corresponding optimal closed-loop diffusion is
\begin{multline}
\label{eq:closed_loop_diffusion}
dX_t
=
\Bigl(
f_0(X_t)
-
B(X_t)\widetilde R_\lambda(X_t)^{-1}B(X_t)^\top
\nabla V^\lambda(X_t)
\Bigr)dt
\\
+
\sigma(X_t)d\mathcal{W}_t.
\end{multline}

Trajectory regularization augments the control weight by the positive semidefinite term
$\lambda B(x)^\top(\sigma(x)\sigma(x)^\top)^{-1}B(x)$.
Thus, drift changes in low-noise directions receive a larger quadratic penalty.
For a fixed value gradient, increasing $\lambda$ decreases $\widetilde R_\lambda^{-1}$ in the positive-semidefinite order; however, $V^\lambda$ also changes with $\lambda$, so no general pointwise monotonicity of $\|u^{*,\lambda}(x)\|$ follows.


\section{Closed-Loop Stability Analysis}
\label{sec:stability}

We study the infinite-horizon closed-loop system under the optimal feedback \eqref{eq:optimal_feedback_stationary}, resulting in the diffusion \eqref{eq:closed_loop_diffusion} for a fixed $\lambda\ge0$.


\subsection{Generator and Foster--Lyapunov Structure}

For brevity, define the closed-loop drift
$b^{*,\lambda}(x)
=
f_0(x)-B(x)\widetilde R_\lambda(x)^{-1}B(x)^\top\nabla V^\lambda(x)$.
Let $\mathcal L^{*,\lambda}$ denote the generator associated with \eqref{eq:closed_loop_diffusion}:
\begin{equation}
\label{eq:generator_closed_loop}
\mathcal L^{*,\lambda}\varphi(x)
=
\nabla \varphi(x)^\top b^{*,\lambda}(x)
+
\frac12
\mathrm{tr}\!\left(\sigma(x)\sigma(x)^\top \nabla^2 \varphi(x)\right),
\end{equation}
for $\varphi\in C^2(\mathbb R^n)$.
We assume that $b^{*,\lambda}$ and $\sigma$ are locally Lipschitz with at most linear growth, so that the closed-loop SDE admits a unique nonexplosive strong solution and defines a Feller Markov process.

\noindent
\textbf{\textit{Assumption 3.}}
There exist a function $W\in C^2(\mathbb R^n;[0,\infty))$ and constants $c>0$, $d\ge0$ such that \cite{meyn2012markov}:
\begin{itemize}
    \item[(i)] $W(x)$ is radially unbounded,
    \item[(ii)] $\mathcal L^{*,\lambda}W(x)\le -cW(x)+d$ for all $x\in\mathbb R^n$.
\end{itemize}

Assumption~3 is a standard Foster--Lyapunov drift condition for boundedness and invariant-measure analysis \cite{meyn2012markov}.
Such a function $W$ arises naturally for dissipative closed-loop dynamics.
Under additional coercivity conditions, the value function itself can also serve as a Lyapunov function.


\subsection{Moment Bounds and Mean-Square Boundedness}

\noindent
\textbf{\textit{Proposition 6 (Exponential Lyapunov Bound).}}
Suppose Assumption~3 holds.
For all $x\in\mathbb R^n$ and all $t\ge0$,
\begin{equation}
\label{eq:exp_moment_bound}
\mathbb E_x[W(X_t)]
\le
e^{-ct}W(x)
+
\frac{d}{c}\bigl(1-e^{-ct}\bigr).
\end{equation}
In particular,
\begin{equation}
\label{eq:uniform_moment_bound}
\sup_{t\ge0}\mathbb E_x[W(X_t)]
\le
W(x)+\frac{d}{c}.
\end{equation}

\noindent
\textit{Proof.}
Applying It\^o's formula to the stopped process gives
\begin{equation}
\label{eq:ito_lyapunov}
W(X_t)
=
W(x)
+
\int_0^t \mathcal L^{*,\lambda}W(X_s)ds
+
M_t,
\end{equation}
where the local martingale term is handled by standard localization.
Taking expectations, removing the localization, and using Assumption~3 yields
\[
\mathbb E_x[W(X_t)]
\le
W(x)
+
\int_0^t
\left(
-c\mathbb E_x[W(X_s)]+d
\right)ds.
\]
Gr\"onwall's inequality gives \eqref{eq:exp_moment_bound}, and \eqref{eq:uniform_moment_bound} follows. $\square$

\noindent
\textbf{\textit{Corollary 1 (Mean-Square Boundedness).}}
Assume, in addition, there exist constants $k_1,k_2>0$ such that
\begin{equation}
\label{eq:quadratic_dominance}
k_1\|x\|^2
\le
W(x)
\le
k_2(1+\|x\|^2)
\quad
\text{for all } x\in\mathbb R^n.
\end{equation}
Then, $\sup_{t\ge0}\mathbb E_x\|X_t\|^2<\infty$, i.e., the second moment of the state is uniformly bounded.

\noindent
\textit{Proof.}
Combine \eqref{eq:uniform_moment_bound} with the lower bound in \eqref{eq:quadratic_dominance}. $\square$


\subsection{Invariant Measures}

We next establish existence of an invariant measure from the Foster--Lyapunov condition.

\noindent
\textbf{\textit{Proposition 7 (Existence of an Invariant Measure).}}
Suppose Assumption~3 holds.
Then, \eqref{eq:closed_loop_diffusion} admits at least one invariant probability measure $\pi$, i.e., $X_0\sim\pi\Rightarrow X_t\sim\pi$.
Moreover,
\begin{equation}
\label{eq:invariant_moment_bound}
\int_{\mathbb R^n} W(x)\pi(dx)<\infty,
\end{equation}
and, for every $\varphi\in C_c^2(\mathbb R^n)$,
\begin{equation}
\label{eq:generator_invariant_identity}
\int_{\mathbb R^n}\mathcal L^{*,\lambda}\varphi(x)\pi(dx)=0.
\end{equation}

\noindent
\textit{Proof.}
For an initial state $x$, define the averaged transition measure
$\bar\mu_T(\cdot)=\frac{1}{T}\int_0^T\mathbb P_x(X_t\in\cdot)\,dt$.
By \eqref{eq:uniform_moment_bound} and radial unboundedness of $W$, the family $\{\bar\mu_T\}_{T>0}$ is tight \cite{meyn2012markov}.
Hence, there exists a weakly convergent subsequence.
Since the closed-loop process is Feller, the Krylov--Bogoliubov theorem implies that every such weak limit is invariant \cite{kryloff1937theorie}.
The integrability condition \eqref{eq:invariant_moment_bound} follows from the uniform $W$-moment bound, and \eqref{eq:generator_invariant_identity} follows from invariance. $\square$



\subsection{Value Function as a Lyapunov Candidate}

Suppose that $V^\lambda\in C^2(\mathbb R^n)$ so that the stationary HJB equation holds.
Then, under the optimal feedback \eqref{eq:optimal_feedback_stationary},
\begin{equation}
\label{eq:value_generator_identity}
\rho V^\lambda(x)
=
L(x,u^{*,\lambda}(x))
+
\mathcal L^{*,\lambda}V^\lambda(x).
\end{equation}
This identity follows directly from the stationary HJB equation \eqref{eq:hjb_stationary} evaluated at the minimizing control.
It provides a direct route to a Foster--Lyapunov inequality when the running cost sufficiently dominates the discounted value.

\noindent
\textbf{\textit{Assumption 4.}}
The value function $V^\lambda$ is radially unbounded, and there exist constants $\eta>0$ and $d\ge0$ such that
$L(x,u^{*,\lambda}(x)) \ge (\rho+\eta)V^\lambda(x)-d$ for all $x\in\mathbb R^n$.

\noindent
\textbf{\textit{Proposition 8 (Value Function as a Foster--Lyapunov Function).}}
Suppose Assumption~4 holds.
Then $V^\lambda(x)$ satisfies
\begin{equation}
\label{eq:value_lyapunov}
\mathcal L^{*,\lambda}V^\lambda(x)
\le
-\eta V^\lambda(x)+d
\quad
\text{for all } x\in\mathbb R^n.
\end{equation}
Hence, since $V^\lambda\ge0$ under the nonnegative stationary running cost, $V^\lambda$ satisfies Assumption~3 with $W=V^\lambda$ and $c=\eta$.

\noindent
\textit{Proof.}
By rearranging \eqref{eq:value_generator_identity},
$\mathcal L^{*,\lambda}V^\lambda(x)
=
\rho V^\lambda(x)-L(x,u^{*,\lambda}(x))$.
Assumption~4 therefore gives
$\mathcal L^{*,\lambda}V^\lambda(x)
\le
\rho V^\lambda(x)-\bigl((\rho+\eta)V^\lambda(x)-d\bigr)
=
-\eta V^\lambda(x)+d$.
$\square$

Thus, when the optimal running cost dominates the discounted value sufficiently strongly in the tails, the value function itself provides a Foster--Lyapunov certificate for the optimal closed-loop diffusion.
The parameter $\lambda$ changes both $\widetilde R_\lambda$ and $V^\lambda$, and therefore changes the closed-loop drift $b^{*,\lambda}$.
Its effect on dissipativity and stability is system-dependent; no monotonic improvement in closed-loop stability with increasing $\lambda$ is implied in general.


\section{Linear--Quadratic Specialization}
\label{sec:lq}

We specialize TRSOC to the LQ setting to show how trajectory regularization modifies the classical discounted LQR problem and its Riccati characterization.


\subsection{Linear Controlled and Reference Diffusions}

Consider linear controlled and reference diffusions
\begin{align}
\label{eq:lq_controlled}
dX_t = (A X_t + B u_t)dt + \Sigma d\mathcal{W}_t, \\
\label{eq:lq_reference}
dX_t = A X_t dt + \Sigma d\mathcal{W}_t,
\end{align}
where $A\in\mathbb R^{n\times n}$, $B\in\mathbb R^{n\times m}$, and $\Sigma\in\mathbb R^{n\times n}$ are constant.
As assumed in the nonlinear case, the diffusion matrix satisfies $\Sigma\Sigma^\top\succ0$.
Then, Lemma~1 gives
\begin{equation}
\label{eq:lq_kl_identity}
\mathrm{KL}(\mathbb P^u\|\mathbb P^0)
=
\frac12
\mathbb E
\left[
\int_0^T
\|\Sigma^{-1}Bu_t\|^2dt
\right].
\end{equation}
Equivalently, $\|\Sigma^{-1}Bu\|^2 = u^\top B^\top(\Sigma\Sigma^\top)^{-1}Bu$.
Thus, in the linear setting, the trajectory regularization becomes an additional quadratic control penalty.


\subsection{Discounted Infinite-Horizon Objective}

Let the running cost be
\begin{equation}
\label{eq:lq_running_cost}
\ell(x,u)
=
x^\top Qx+\frac12 u^\top Ru,
\qquad
Q\succeq0,
\quad
R\succ0,
\end{equation}
and consider the discounted infinite-horizon objective
\begin{multline}
\label{eq:lq_objective}
J_x^\lambda(u)
=
\mathbb E_x
\biggl[
\int_0^\infty
e^{-\rho t}
\Bigl(
X_t^\top QX_t
+
\frac12 u_t^\top Ru_t
\\
+
\frac{\lambda}{2}
u_t^\top B^\top(\Sigma\Sigma^\top)^{-1}Bu_t
\Bigr)dt
\biggr].
\end{multline}

In this case, the effective control weight \eqref{eq:effective_control_weight} reduces to
\begin{equation}
\label{eq:lq_effective_control_weight}
\widetilde R_\lambda
=
R+\lambda B^\top(\Sigma\Sigma^\top)^{-1}B,
\end{equation}
which is positive definite.
Hence
\begin{equation}
\label{eq:lq_objective_reduced}
J_x^\lambda(u)
=
\mathbb E_x
\left[
\int_0^\infty
e^{-\rho t}
\left(
X_t^\top QX_t
+
\frac12 u_t^\top \widetilde R_\lambda u_t
\right)dt
\right].
\end{equation}
Therefore, in the linear special case, TRSOC reduces to a discounted LQR problem with modified control weight $\widetilde R_\lambda$.


\subsection{Quadratic Value Function and Riccati Equation}

We seek a quadratic value function of the form
\begin{equation}
\label{eq:lq_value_function}
V^\lambda(x)
=
\frac12 x^\top P_\lambda x + c_\lambda,
\qquad
P_\lambda=P_\lambda^\top.
\end{equation}
Then $\nabla V^\lambda(x)=P_\lambda x$ and $\nabla^2V^\lambda(x)=P_\lambda$.
Substituting \eqref{eq:lq_value_function} into the stationary HJB equation \eqref{eq:hjb_reduced} yields
\begin{multline}
\label{eq:lq_hjb}
\rho\left(\frac12 x^\top P_\lambda x + c_\lambda\right)
=
x^\top Qx
+
x^\top P_\lambda Ax
+
\frac12 \mathrm{tr}(\Sigma\Sigma^\top P_\lambda)
\\
-
\frac12 x^\top P_\lambda B \widetilde R_\lambda^{-1} B^\top P_\lambda x.
\end{multline}
The optimal feedback law is
\begin{equation}
\label{eq:lq_optimal_feedback}
u^{*,\lambda}(x)
=
-\widetilde R_\lambda^{-1}B^\top P_\lambda x,
\end{equation}
which is the linear specialization of \eqref{eq:optimal_feedback_stationary}.

Matching the quadratic terms in \eqref{eq:lq_hjb} gives the discounted algebraic Riccati equation (ARE) \cite{bertsekas2012dynamic}
\begin{equation}
\label{eq:lq_are}
\rho P_\lambda
=
2Q + A^\top P_\lambda + P_\lambda A
-
P_\lambda B \widetilde R_\lambda^{-1} B^\top P_\lambda,
\end{equation}
and matching the constant terms gives
\begin{equation}
\label{eq:lq_constant_term}
\rho c_\lambda
=
\frac12 \mathrm{tr}(\Sigma\Sigma^\top P_\lambda),
\quad
c_\lambda
=
\frac{1}{2\rho}\mathrm{tr}(\Sigma\Sigma^\top P_\lambda).
\end{equation}

\subsection{Existence of a Stabilizing Solution}

Define the discount-shifted matrix $\bar A:=A-\frac{\rho}{2}I$.

\noindent
\textbf{\textit{Proposition 9 (Discounted ARE and Closed-Loop Stability).}}
Suppose $(\bar A,B)$ is stabilizable and $(Q^{1/2},\bar A)$ is detectable.
Then, \eqref{eq:lq_are} admits a unique stabilizing solution $P_\lambda\succeq0$, and
$\bar A-B\widetilde R_\lambda^{-1}B^\top P_\lambda$ is Hurwitz.
The physical closed-loop matrix is
\begin{equation}
\label{eq:lq_closed_loop_matrix}
A_{\mathrm{cl},\lambda}
=
A-B\widetilde R_\lambda^{-1}B^\top P_\lambda.
\end{equation}
Discounted stability does not by itself imply that $A_{\mathrm{cl},\lambda}$ is Hurwitz.
If $A_{\mathrm{cl},\lambda}$ is Hurwitz, then the closed-loop diffusion
\begin{equation}
\label{eq:lq_closed_loop_diffusion}
dX_t
=
A_{\mathrm{cl},\lambda}X_t dt
+
\Sigma d\mathcal{W}_t
\end{equation}
is mean-square bounded and admits a unique Gaussian invariant measure.

\noindent
\textit{Proof.}
Equation~\eqref{eq:lq_are} is the standard continuous-time ARE associated with the shifted matrix $\bar A=A-\rho I/2$ and control weight $\widetilde R_\lambda$.
The invariant-measure claim follows from standard Ornstein--Uhlenbeck theory when the unshifted matrix $A_{\mathrm{cl},\lambda}$ is Hurwitz. $\square$

In the LQ case, trajectory regularization modifies the classical discounted LQR problem through the matrix $\widetilde R_\lambda$.
When $\lambda=0$, the standard discounted LQR problem is recovered.
Increasing $\lambda$ increases the quadratic penalty on drift changes, particularly in low-noise directions through $(\Sigma\Sigma^\top)^{-1}$.
However, because $P_\lambda$ also depends on $\lambda$, neither the optimal feedback gain nor physical closed-loop stability is generally monotone in $\lambda$.


\begin{figure}[t]
    \centering
    \begin{subfigure}[t]{0.49\linewidth}
        \centering
        \includegraphics[width=\linewidth]{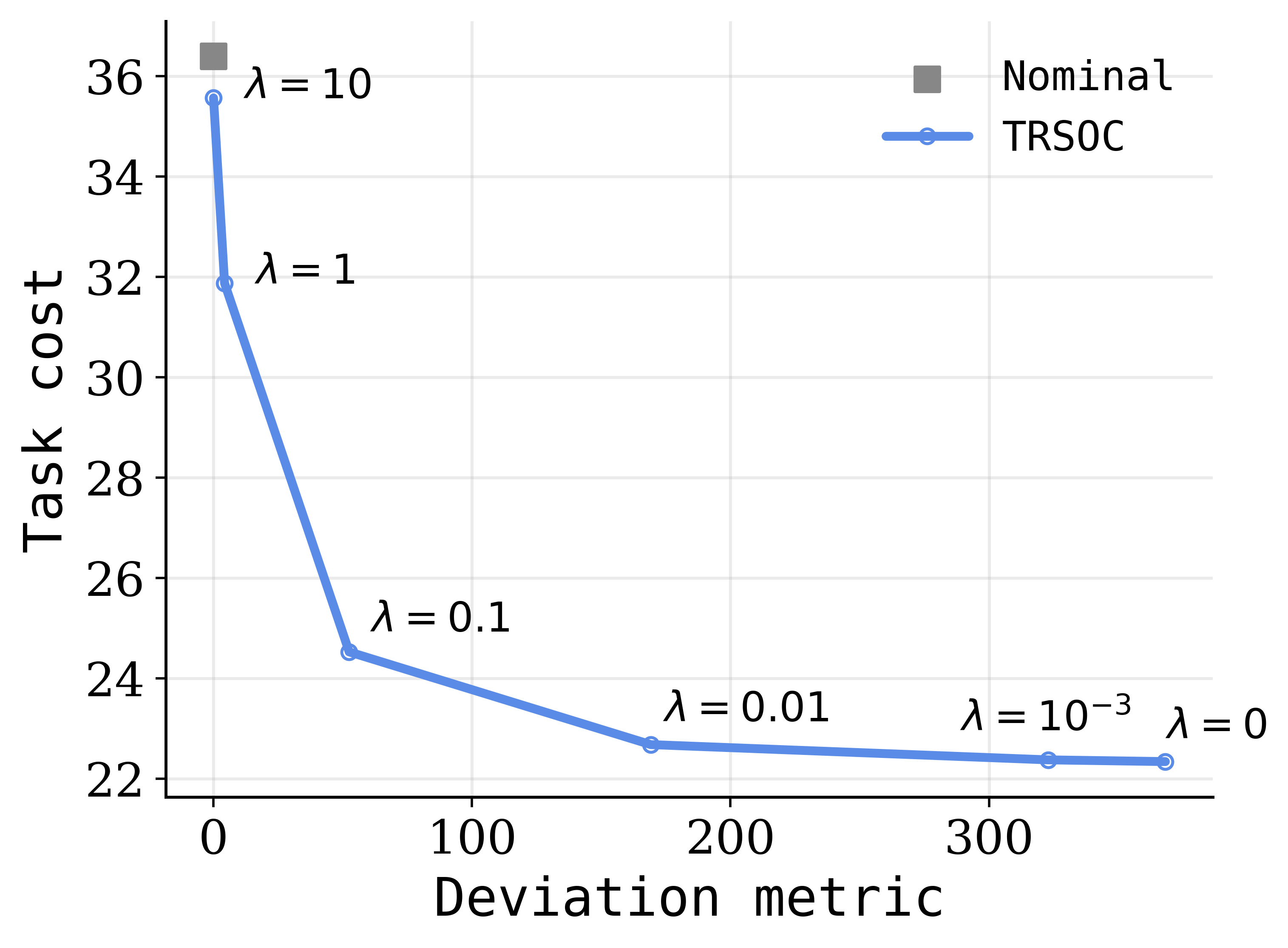}
        \caption{Performance--deviation trade-off with known reference control.}
        \label{fig:num_pareto_known}
    \end{subfigure}
    \hfill
    \begin{subfigure}[t]{0.49\linewidth}
        \centering
        \includegraphics[width=\linewidth]{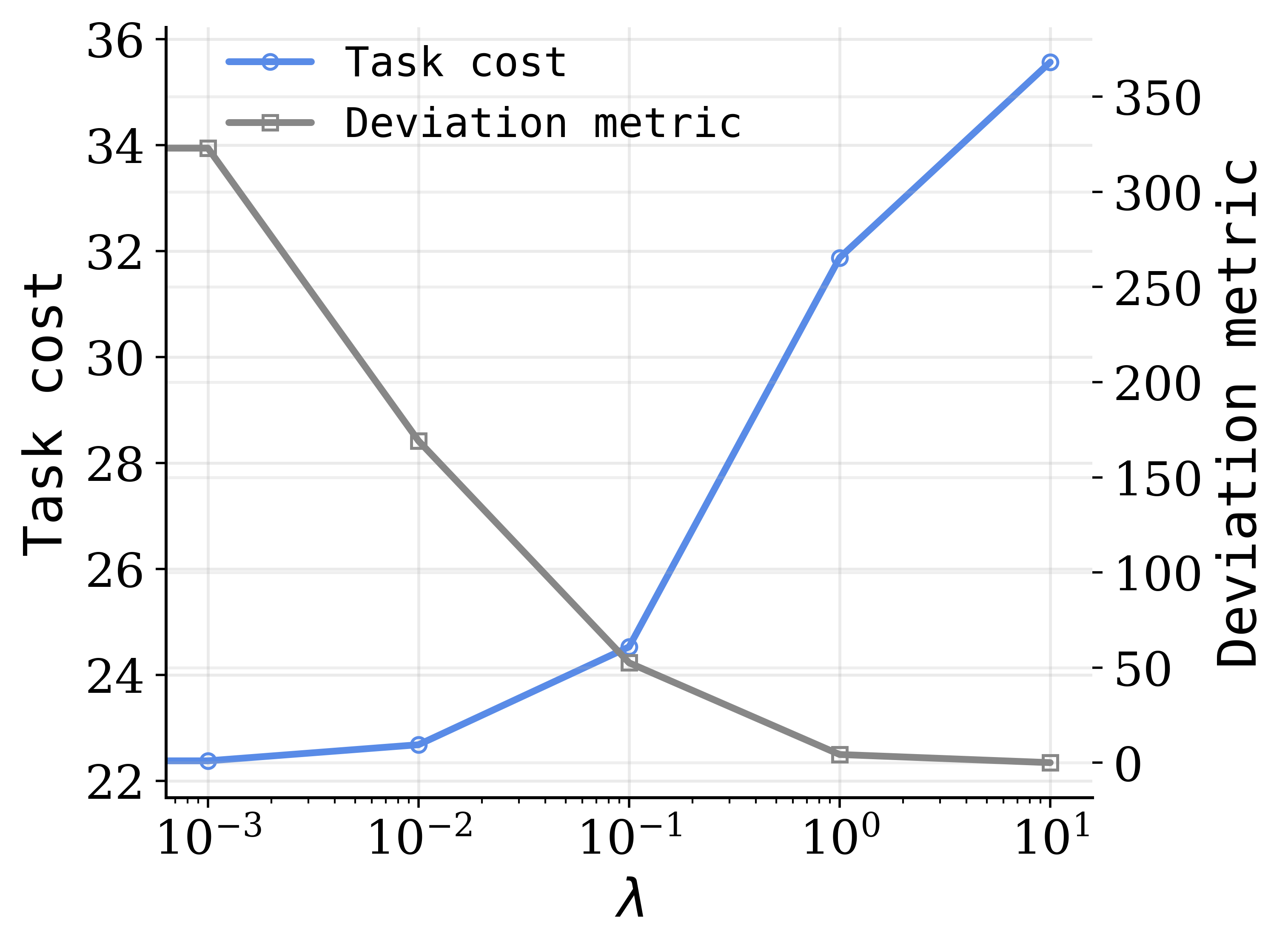}
        \caption{Cost and deviation metric as functions of $\lambda$.}
        \label{fig:num_lambda_sweep_known}
    \end{subfigure}
    \caption{
    \textbf{Known-reference TRSOC.}
    The left panel shows the trade-off between task performance and deviation from the nominal reference.
    The right panel shows the same trend as a function of $\lambda$.
    As $\lambda$ increases, the controller shifts from performance-driven to reference-preserving behavior.
    }
    \label{fig:num_known_trade-off}
    \vspace{-1em}
\end{figure}


\section{Numerical Results}
\label{sec:results}

We illustrate the behavior induced by TRSOC through tracking problems based on discretized controlled diffusions.
Rather than improving over the unregularized SOC optimum, the experiments visualize the trade-off between task performance and deviation from a reference behavior.

For known references, the finite-horizon quadratic problem is solved by backward DP.
For data-driven references, the reference dynamics are learned from offline trajectories, and TRSOC is solved over time-varying affine feedback policies.


\subsection{Experimental Setup}

We consider a two-dimensional stochastic double-integrator with state
$x_t = [p_{x,t}, p_{y,t}, v_{x,t}, v_{y,t}] \in \mathbb R^4$
and control
$u_t = [u_{x,t}, u_{y,t}] \in \mathbb R^2$.
The system dynamics are
\begin{equation}
\label{eq:num_sde}
dp_t = v_t\,dt + \Sigma_p\,dW_t^{(p)},
\quad
dv_t = u_t\,dt + \Sigma_v\,dW_t^{(v)},
\end{equation}
where $W_t^{(p)}$ and $W_t^{(v)}$ are independent Wiener processes.
We assume $\Sigma_p$ and $\Sigma_v$ are nonsingular, so the full diffusion is consistent with Assumption~1.
In all experiments, we discretize with time step $\Delta t = 0.05$ and evaluate the controllers on the stochastic system.

The tracking target is a figure-eight trajectory
\begin{equation}
\label{eq:num_fig8}
p_{\mathrm{ref}}(t)
=
\begin{bmatrix}
a\sin(\omega t)\\
b\sin(\omega t)\cos(\omega t)
\end{bmatrix},
\end{equation}
with corresponding reference velocity $v_{\mathrm{ref}}(t)=\dot p_{\mathrm{ref}}(t)$ and feedforward acceleration $u_{\mathrm{ff}}(t)=\ddot p_{\mathrm{ref}}(t)$.
The task objective is the finite-horizon tracking cost
\begin{equation}
\label{eq:num_task_cost}
J_{\mathrm{task}}(u)
=
\mathbb E
\left[
\int_0^T
\Bigl(
(x_t-x_{\mathrm{ref}}(t))^\top Q (x_t-x_{\mathrm{ref}}(t))
+
u_t^\top R u_t
\Bigr)dt
\right],
\end{equation}
where $x_{\mathrm{ref}}(t)=[p_{\mathrm{ref}}(t),v_{\mathrm{ref}}(t)]$ and $Q\succeq0$, $R\succ0$.

For a reference controller $u^0$, the controlled and reference drifts differ only through the acceleration channel.
We therefore use the corresponding drift-energy metric
\begin{equation}
\label{eq:num_dev_metric}
D(u)
=
\mathbb E
\left[
\int_0^T
\left\|
\Sigma_v^{-1}\bigl(u_t-u_t^{\mathrm{ref}}\bigr)
\right\|^2 dt
\right],
\end{equation}
which satisfies $D(u)=2\,\mathrm{KL}(\mathbb P^u\|\mathbb P^0)$ by Lemma~1.
In the data-driven experiment, $u_t^{\mathrm{ref}}$ is replaced by the learned reference acceleration $\hat a_0(t,x_t)$, with learned drift $\hat f_0(t,x)=[v,\hat a_0(t,x)]$.
Thus, the deviation metric becomes
\begin{equation}
\label{eq:num_dev_metric_data}
D_{\mathrm{data}}(u)
=
\mathbb E
\left[
\int_0^T
\left\|
\Sigma_v^{-1}\bigl(u_t-\hat a_0(t,x_t)\bigr)
\right\|^2 dt
\right].
\end{equation}

All costs and deviation metrics are estimated by Monte--Carlo rollouts under the stochastic dynamics.
For each controller, we report mean values over multiple noise realizations.


\begin{figure}[t]
    \centering
    \begin{subfigure}[t]{0.49\linewidth}
        \centering
        \includegraphics[width=\linewidth]{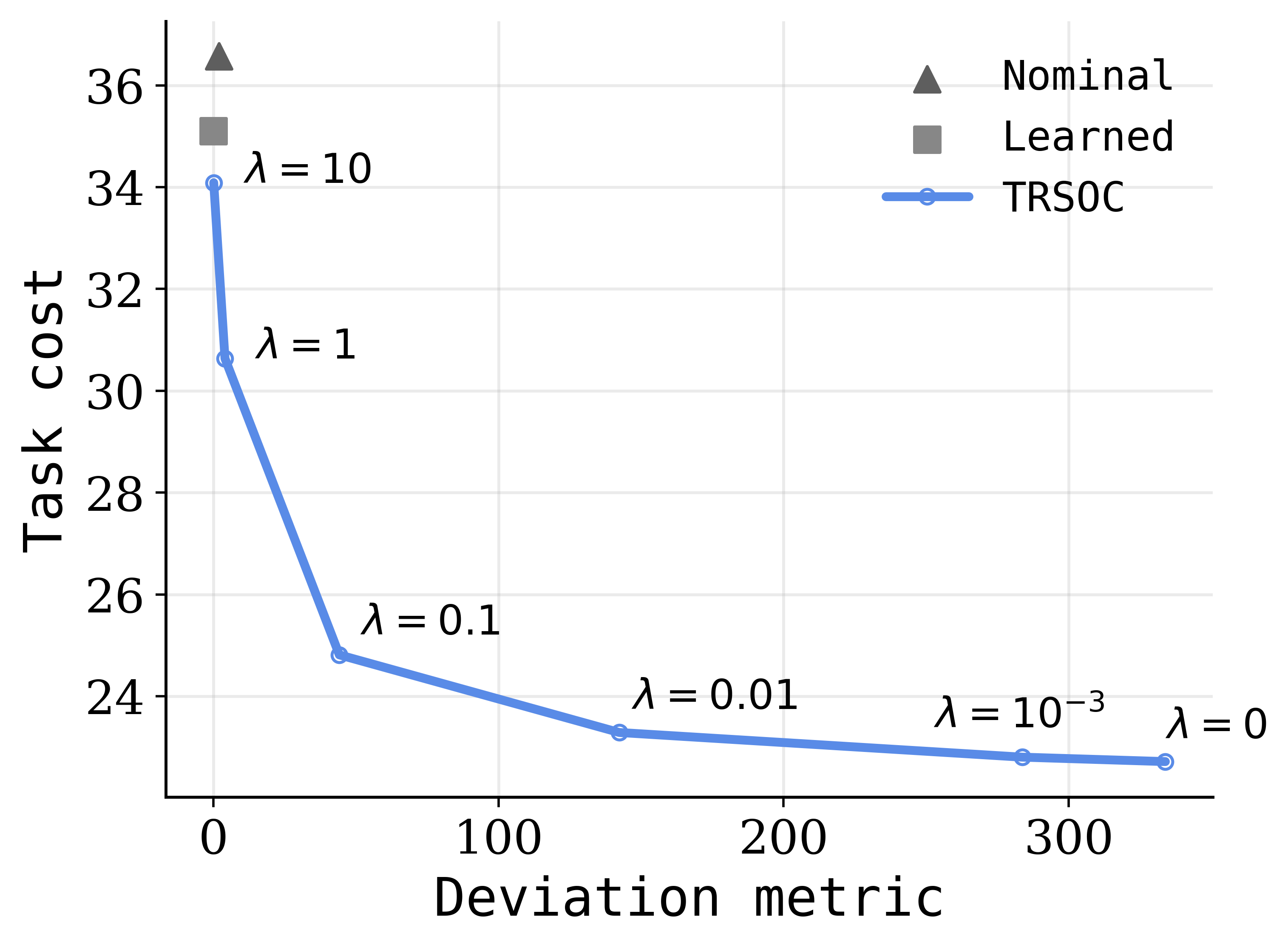}
        \caption{Performance--deviation trade-off with learned reference.}
        \label{fig:num_pareto_data}
    \end{subfigure}
    \hfill
    \begin{subfigure}[t]{0.49\linewidth}
        \centering
        \includegraphics[width=\linewidth]{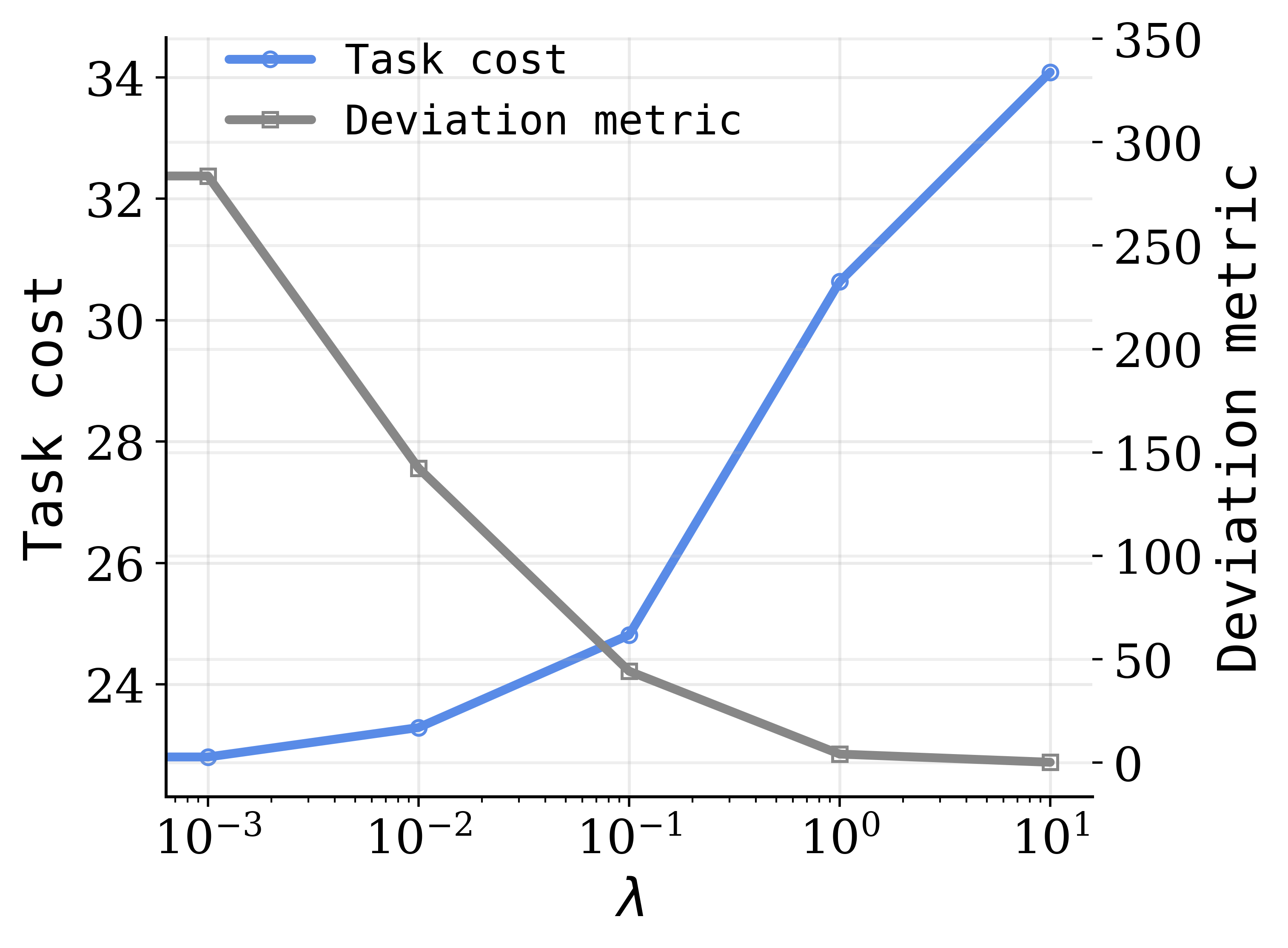}
        \caption{Cost and deviation from learned reference w.r.t. $\lambda$.}
        \label{fig:num_lambda_sweep_data}
    \end{subfigure}
    \caption{
    \textbf{Data-driven TRSOC with learned reference dynamics.}
    The left panel shows the trade-off between task performance and deviation from the learned reference.
    The right panel shows the same trend as a function of $\lambda$.
    }
    \label{fig:num_data_trade-off}
    \vspace{-1em}
\end{figure}


\subsection{Known Reference Tradeoff Across $\lambda$}

We consider the case where reference behavior is given by a known nominal tracking controller.
Specifically, we define
\begin{equation}
\label{eq:num_nominal_ctrl}
u^0(t,x)
=
u_{\mathrm{ff}}(t)
-
K_p\bigl(p-p_{\mathrm{ref}}(t)\bigr)
-
K_d\bigl(v-v_{\mathrm{ref}}(t)\bigr),
\end{equation}
and use the induced drift as the reference dynamics.
The finite-horizon TRSOC problem is
\begin{equation}
\label{eq:num_known_ref_obj}
J_{\mathrm{TRSOC}}^\lambda(u)
=
J_{\mathrm{task}}(u)
+
\frac{\lambda}{2}
D(u).
\end{equation}
By Lemma~1, the second term is exactly $\lambda\mathrm{KL}(\mathbb P^u\|\mathbb P^0)$.
The discretized problem remains quadratic and is solved exactly by backward DP.

Figure~\ref{fig:num_pareto_known} shows the trade-off between task performance and deviation from the nominal behavior \eqref{eq:num_nominal_ctrl} as $\lambda$ varies.
The unregularized controller ($\lambda=0$) achieves the lowest task cost but significantly deviates from the nominal behavior.
As $\lambda$ increases, deviation decreases while task cost increases, with the nominal controller near the zero-deviation end.
Figure~\ref{fig:num_lambda_sweep_known} shows the same transition from performance-driven to reference-preserving behavior.


\begin{figure}[t]
    \centering
    \begin{subfigure}[t]{0.49\linewidth}
        \centering
        \includegraphics[width=\linewidth]{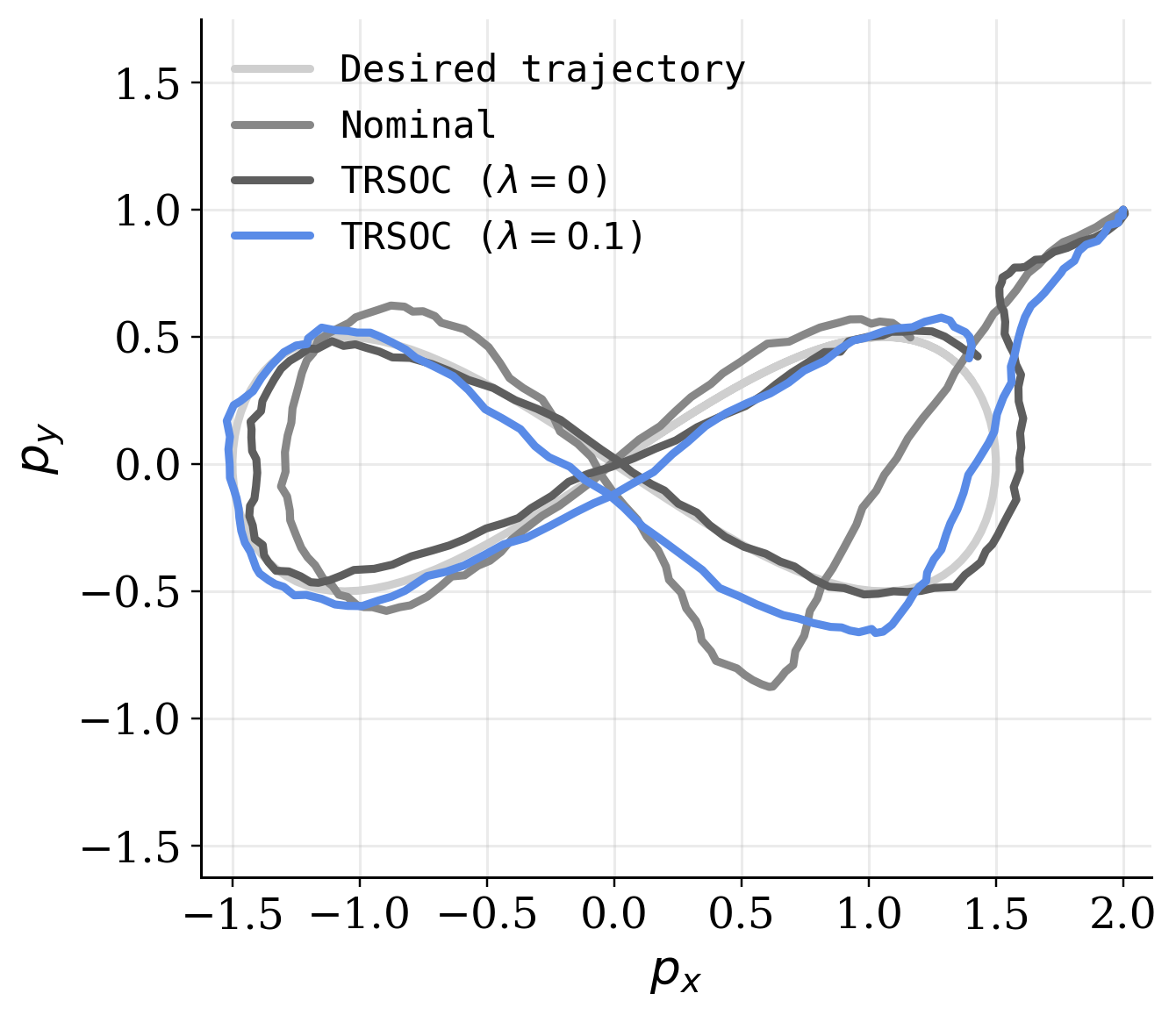}
        \caption{Known reference setting.}
        \label{fig:num_traj_known}
    \end{subfigure}
    \hfill
    \begin{subfigure}[t]{0.49\linewidth}
        \centering
        \includegraphics[width=\linewidth]{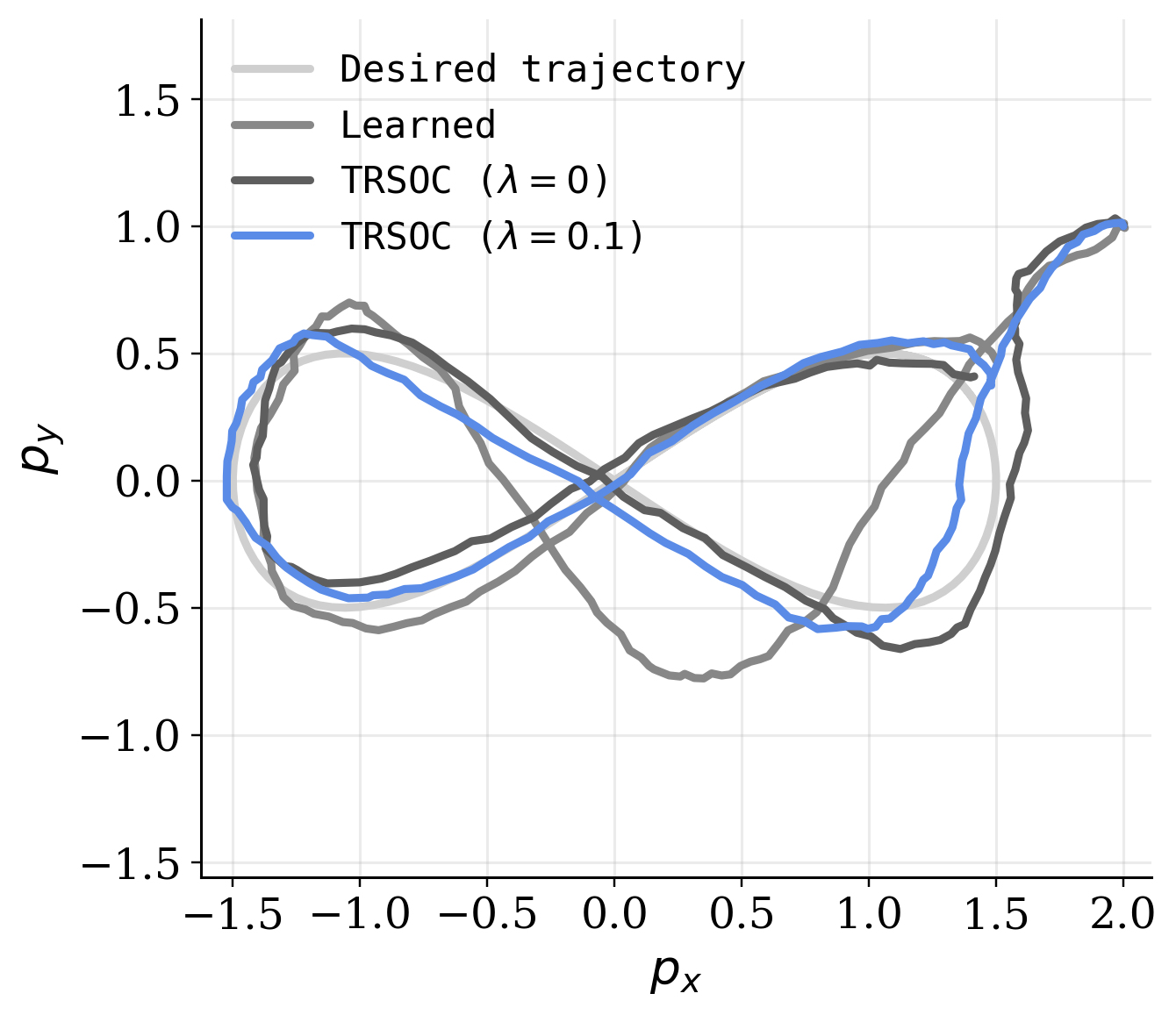}
        \caption{Learned reference setting.}
        \label{fig:num_traj_data}
    \end{subfigure}
    \caption{
    \textbf{Trajectories induced by TRSOC.}
    Rollouts are shown for the reference behavior, the unregularized controller ($\lambda=0$), and a regularized controller with intermediate $\lambda$.
    Increasing $\lambda$ shifts the induced state trajectory toward the corresponding reference behavior in both settings.
    }
    \label{fig:num_traj_both}
    \vspace{-1em}
\end{figure}


\subsection{Data-Driven TRSOC with Learned Reference Dynamics}

We next consider data-driven TRSOC where the reference behavior is learned from offline trajectories rather than given analytically.
This setting is common in both control and RL, where one seeks to preserve nominal behavior while optimizing a task cost.
We collect offline data from rollouts of the nominal controller \eqref{eq:num_nominal_ctrl}.
For the double integrator, the position drift $\dot p=v$ is known, so only the acceleration must be learned.
We train a small neural network to approximate the acceleration $\hat a_0(t,x)$ and define $\hat f_0(t,x)=[v,\hat a_0(t,x)]$.
The resulting data-driven TRSOC objective is
\begin{equation}
\label{eq:num_data_driven_obj}
J_{\mathrm{TRSOC}}^\lambda(u)
=
J_{\mathrm{task}}(u)
+
\frac{\lambda}{2}
D_{\mathrm{data}}(u).
\end{equation}
Since the learned reference drift breaks the LQ structure, we solve numerically over time-varying affine feedback policies.

\begin{figure}[t]
    \centering
    \includegraphics[width=0.95\linewidth]{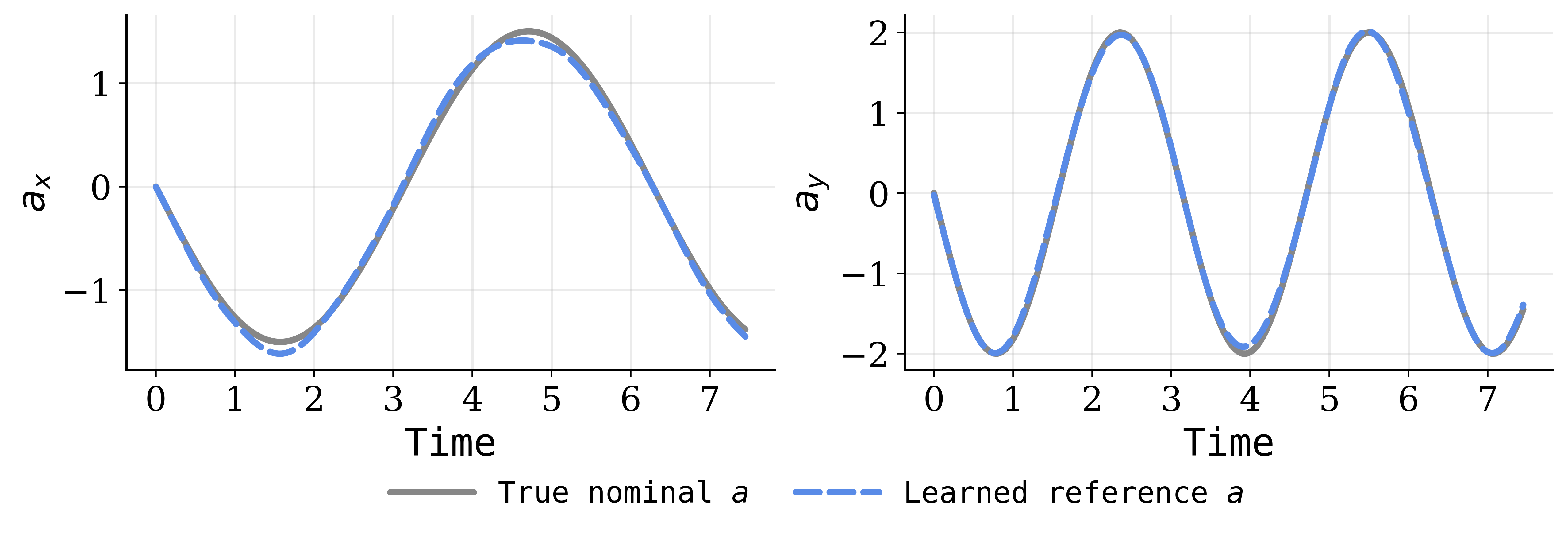}
    \caption{
    \textbf{Approximation quality of the learned reference drift.}
    The learned acceleration $\hat a_0(t,x)$ is compared with the nominal acceleration along the reference trajectory.
    The close agreement shows that the learned drift captures the nominal behavior used in the data-driven TRSOC.
    }
    \label{fig:num_fit_data}
    \vspace{-1em}
\end{figure}

Figure~\ref{fig:num_fit_data} shows that the learned drift closely approximates the nominal acceleration along the reference trajectory.
Figures~\ref{fig:num_pareto_data} and \ref{fig:num_traj_data} show that the same TRSOC trade-off persists in the data-driven setting.
As $\lambda$ increases, the controller shifts from performance-driven behavior toward the learned reference.
This demonstrates the applicability of TRSOC to reference dynamics estimated from offline data.


\subsection{Trajectory Visualization}

We visualize the trajectory-level effect of the regularization.
Figure~\ref{fig:num_traj_known} compares representative rollouts of the nominal controller, the unregularized controller ($\lambda=0$), and a TRSOC controller with intermediate $\lambda$.
Because the nominal controller contains proportional feedback, it reacts strongly to the initial tracking error and quickly approaches the reference curve.
The regularized controller exhibits a less aggressive initial response and subsequently approaches the task-driven trajectory.
More generally, TRSOC produces trajectories between the nominal and unregularized behaviors.
Larger $\lambda$ keeps the state evolution closer to the reference, particularly in regions where the task-optimal corrections differ from the nominal behavior.
Thus, trajectory regularization directly changes the induced state trajectories rather than merely increasing the scalar control cost.


\section{Conclusion}

We proposed TRSOC, which augments SOC with trajectory-level KL regularization.
Using Girsanov's theorem, we showed that the finite-horizon trajectory KL reduces exactly to a quadratic drift-mismatch penalty while preserving the DP and HJB structure.
The framework provides a tunable trade-off between performance-driven and reference-preserving behavior through the regularization parameter.
In the LQ setting, TRSOC admits an explicit Riccati characterization with a modified, noise-aware control cost.
Numerical results demonstrate the predicted trade-off at both the objective and trajectory levels.
The framework also applies when the reference dynamics are estimated from offline data.


\addtolength{\textheight}{-12cm}


\printbibliography
\end{document}